\documentclass{aa}
\usepackage{graphicx}   
\newdimen\captwidth   
\newdimen\figwidth   
\def\eg{{\it e.g. }}
\def\etal{{et al. }}  
\def\ie{{\it i.e. }}
\def\micron{\hbox{$\,\mu {\rm m}\,$}}
\def\milli{\hbox{$\, {\rm mm}\,$}}
\def\zu{\rm\,}     

\begin{document}
 
   \thesaurus{22         
     (03.09.1;  
      03.09.4;  
      03.13.5;  
      12.03.3;  
      12.03.3;  
      13.18.2)  
            }
%
 
\title{Calibration and First light of the Diabolo photometer at the
  Millimetre and Infrared Testa Grigia Observatory}
 
 
\author{ A. Benoit \inst{1} \and F. Zagury \inst{2} \and N. Coron
  \inst{2} \and M. De Petris \inst{3} F.-X. D\'esert \inst{4} \and M.
  Giard \inst{5} \and J.--P. Bernard \inst{2} \and J.--P. Crussaire
  \inst{2} \and G. Dambier \inst{2} \and P. de Bernardis \inst{3} \and
  J. Delabrouille \inst{2,10} \and A. De Luca \inst{2} \and P. de
  Marcillac \inst{2} \and G. Jegoudez \inst{2} \and J.--M. Lamarre
  \inst{2} \and J. Leblanc \inst{2} \and J.--P. Lepeltier \inst{2}
  \and B. Leriche \inst{2} \and G. Mainella \inst{3,9} 
  \and J. Narbonne \inst{5} \and F. Pajot
  \inst{2} \and R. Pons \inst{5} J.-L. Puget \inst{2} \and S. Pujol
  \inst{6} \and G. Recouvreur \inst{2} \and G. Serra \inst{5} \and V.
  Soglasnova \inst{7} \and J.--P. Torre \inst{8} \and B. Vozzi
  \inst{2,3} }
 
   \offprints{F.--X. D\'esert$^4$}
 
   \institute{ 
Centre de Recherche sur les Tr\`es Basses Temp\'eratures,
25 Avenue des Martyrs BP166, F--38042 Grenoble Cedex 9, France
\and
Institut d'Astrophysique Spatiale, B\^at. 121,
Universit\'e Paris XI, F--91405 Orsay Cedex, France 
\and
Gruppo di Cosmologia Sperimentale, Dipartimento di Fisica, Universita
``La Sapienza'', P. A. Moro, 2, 00185 Roma, Italia
\and
Laboratoire d'Astrophysique de 
l'Observatoire de Grenoble, 414 rue de la Piscine, BP53, 
F--38041 Grenoble Cedex 9, France
\and
Centre d'\'Etude Spatiale des Rayonnements, 9 avenue du
Colonel Roche, BP 4346, F--31029 Toulouse Cedex France,
\and
Institut Laue Langevin, avenue des Martyrs, F--38042 Grenoble, France
\and
Space Research Institute, Astrospace Center, Academy of
Science of Russia, Profsoyuznaja St. 84/32 117810 Moscow Russia,
\and
Service d'A\'eronomie, BP 3, F--91371 Verrieres--Le--Buisson,  France
\and
Telescopio THEMIS, c/o Instituto de Astrofisica de Canarias, 
calle Via Lactea, s/n, 38250 La Laguna, Tenerife  -  Spain
\and
Laboratoire de Physique Corpusculaire et Cosmologie, College de
France,  11 pl. Marcelin Berthelot, F-75231 Paris Cedex 5
}
 
\date{Publisher version Fri, 26 Nov 1999} 
\authorrunning{Benoit, \etal}
\titlerunning{The Diabolo photometer at the
  Millimetre and Infrared Testa Grigia Observatory}
   \maketitle
   
\newcommand{\ps}{{\sc Planck}}
\newcommand{\pl}{{\sc Planck}}

   \begin{abstract} We have designed and built a large--throughput dual
     channel photometer, Diabolo. This photometer is dedicated to the
     observation of millimetre continuum diffuse sources, and in
     particular, of the Sunyaev--Zel'dovich effect and of anisotropies
     of the 3K background. We describe the optical layout and
     filtering system of the instrument, which uses two bolometric
     detectors for simultaneous observations in two frequency channels
     at $1.2$ and $2.1 \milli$.  The bolometers are cooled to a
     working temperature of $0.1 \zu K$ provided by a compact dilution
     cryostat.
     The photometric and angular responses of the instrument are
     measured in the laboratory. First astronomical light was detected
     in March 1995 at the focus of the new Millimetre and Infrared
     Testa Grigia Observatory (MITO) Telescope. The established
     sensitivity of the system is of $7 \zu mK_{RJ} s^{1/2}$.  For a
     typical map of at least 10 beams, with one hour of integration
     per beam, one can achieve the rms values of $y_{SZ} \simeq
     7\times 10^{-5}$ and the 3K background anisotropy ${\Delta T\over
       T} \simeq 7\times 10^{-5}$, in winter conditions. We also
     report on a novel bolometer AC readout circuit which allows for
     the first time total power measurements on the sky.  This
     technique alleviates (but does not forbid) the use of chopping
     with a secondary mirror. This technique and the dilution fridge
     concept will be used in future scan--modulated space instrument
     like the ESA \ps\ mission project.

 
   \end{abstract}
 
\section{Introduction}
\label{se:in}

The continuum emission of various astrophysical objects in the
millimetre domain has long been proposed as one important clue to many
physical processes in the Universe: such emission includes dust,
free--free, synchrotron emissions, but also fluctuations of the Cosmic
Microwave Background (CMB), either primordial (Smoot \etal 1992) or
due to intervening matter (Sunyaev \& Zel'dovich 1972).  In the past
15 years, the field of millimetre and far infrared measurements has
tremendously grown.  The advances in instrument technology have
allowed many discoveries, with ground-based observations of our Galaxy
and of extragalactic sources, with the many successful ground-based
and balloon-borne CMB anisotropy experiments, and with the instruments
onboard the {\sc COBE} satellite.
Following the experience acquired with the submillimetre
balloon--borne {\sc PRONAOS--SPM} experiment (Lamarre \etal 1994), we have
devised a millimetre photometer called Diabolo, with two channels
matching the relatively transparent atmospheric spectral windows
around 1.2 and 2.1 \milli.  This instrument is designed to be used for
ground--based observations, taking advantage of the large area
provided by millimetre antennas such as  the 30~m telescope of IRAM,
and of long integration times that can be obtained on a small
dedicated telescope. Such observations are complementary to those that
can be made with highly-performing but costly and resolution--limited
space--borne instruments or short duration balloon--borne experiments.

There are two main disadvantages to ground--based measurements, which are:
\begin{itemize} 
\item a larger background, which not only produces a larger photon
  noise but also limits the sensitivity of bolometers because of their
  power load, especially when one tries to obtain broad--band
  measurements with a throughput ($A\Omega$) much larger than the diffraction
  limit
\item additional sky noise, mainly due to the fluctuating water vapour
  content in the atmosphere, and which is usually the main limitation
  of ground--based instruments unless properly subtracted (see \eg
  Matthews 1980, Church 1995, Melchiorri \etal 1996).
\end{itemize}

There are two usual methods for the subtraction of sky noise, either
spatial or spectral ones.  The spatial subtraction method uses several
detectors in the focal plane of the instrument and takes advantage of
the spatial correlation of the atmospheric noise. For this technique
to work, the source size must be smaller than the array size.
It is especially suited for big telescopes for which the beams from
the different detectors have not diverged much when crossing the 2--3
kilometre high water vapour layer. Kreysa \etal (1990), Wilbanks \etal
(1990), and Gear \etal (1995) have used this technique with bolometer
arrays (MPIfR bolometer arrays, SuZie photometer and SCUBA arrays
respectively).

\begin{figure}[tb]
  \includegraphics[angle=90,width=\columnwidth,origin=br]
    {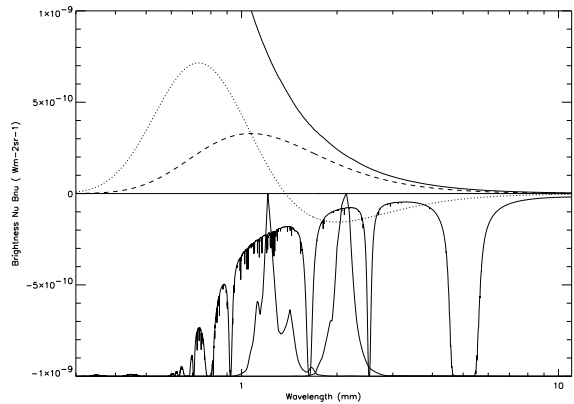}
      \caption{The distortions of the 3K Cosmic Microwave Background
        expected in the millimetre domain are shown in the upper
        panel.  The dotted curve corresponds to the Sunyaev--Zeldovich
        effect with a comptonisation parameter $y=10^{-4}$. The dashed
        curve corresponds to a Doppler effect of ${\Delta T/
          T}=10^{-4}$.  The thin curve shows the spectrum that
        comes from a typical fluctuating part of the atmospheric
        spectrum, normalised to only 0.15\micron of water vapour
        (notice the very different colours between 1 and 2~mm). In
        the lower panel, the curves show the normalised transmission
        of the 2 Diabolo channels centered at 1.2 and 2.1~mm and the
        atmospheric transmission for 3~mm of water vapour (with $y$ axis
        between 0 and 1). Atmospheric curves were deduced from the ATM
        atmospheric model kindly made available by Pardo (1996).}
         \label{SZ}
   \end{figure}

The spectral subtraction method takes advantage of the correlation of
the atmospheric signal at different wavelengths. If the source signal
has a continuum spectrum different from the water vapour emission, one
can form a linear combination of the source fluxes at different
wavelengths which should be quite insensitive to sky noise. This
technique has been used in various photometers. For extended sources
like clusters of galaxies (see below), it has been used by Meyer \etal
(1983), Chase \etal (1987) and Andreani \etal (1996, see also Pizzo
\etal 1995).  In particular the spectra of the 3K CMB distortions
(either primordial or secondary) are quite different from the water
vapour emission as can be seen in Fig~\ref{SZ}. This technique implies
that the smaller wavelength channels do not work at the diffraction
limit, so that the beams at the different wavelengths are
co-extensive. Hence, for broad continuum measurements, the detectors
can only be large-throughput bolometers.

   Once and if the sky noise can be subtracted, the need for sensitive
   large-throughput bolometers implies the lowest possible working
   temperature (see Sect.~\ref{se:dil} \&
   Subsect.~\ref{ss:centmilliK}). Diabolo has been built following
   this line of thought.  It is a simple dual-channel photometer, with
   two bolometers cooled to 0.1~Kelvin for atmospheric noise
   subtraction using the spectral subtraction method adapted to small
   telescopes.  Its design and performance are described in the rest
   of this paper, which is organised as follows. Section~\ref{se:op}
   describes the optical layout of the photometers and the filters we
   use for the proper selection of wavelengths. Section~\ref{se:dil}
   describes the dilution cryostat that is used to cool the
   bolometers.  Section~\ref{se:de} deals with the design and testing
   of the 2 bolometers. Section~\ref{se:re} gives details on the new
   bolometer AC readout electronic circuit which is used for the
   measurements.  Section~\ref{se:fi} gives the characterisation of
   the instrument that was possible with the first observations at the
   new 2.6~metre telescope at Testa Grigia (Italy). Finally, we
   discuss in Section~\ref{se:imp} the recent improvements that have
   been made over the original design.

 
\section{Optical and Filtering systems}
\label{se:op}

\subsection{Optical system}

\begin{figure}[t]
  \includegraphics[angle=0,width=\columnwidth,origin=br]
  {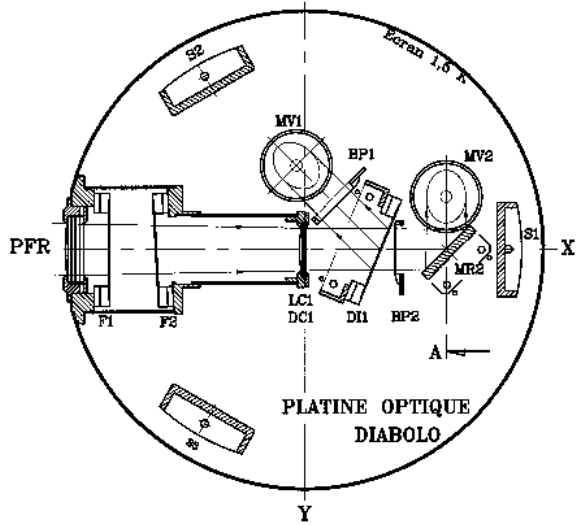}
      \caption{Cold optical plate of the photometer (diameter of
        250~mm): PFR is the cold
        pupil entrance with a lens and some filters, F1, F2 are some
        submillimetre cutoff filters, LC1 is the focal plane lens with
        its diaphragm DC1. A dichroic DI1 splits the radiation between
        channel 1 in reflection and channel 2 in transmission. BP1 and
        BP2 are bandpass filters. MR2 and MV1, MV2 are plane mirrors
        to fold the two beams. S1, S2 and S3 are three pillars holding
        the optical plate to the cryostat.  }
         \label{coldplate}
   \end{figure}

\begin{figure}[t]
  \includegraphics[angle=0,width=\columnwidth,origin=br]
  {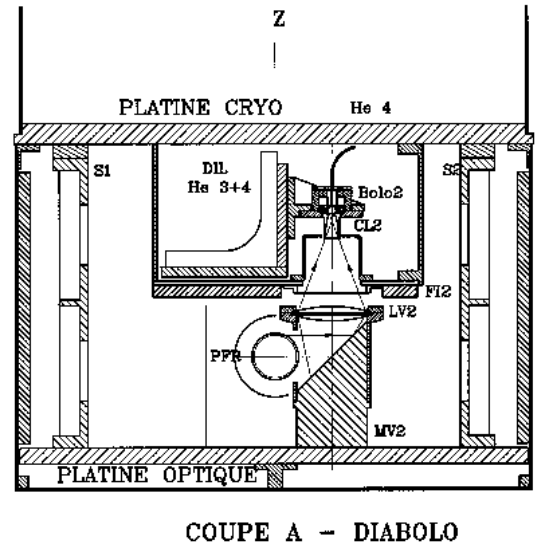}
      \caption{Vertical cut of the cryostat showing the cryogenic
        plate of the photometer and the optical plate. For each
        channel, a lens LV reimages the beam onto the entrance of the
        Winston cone CL, through a bandpass filter FI. The dilution
        fridge provides cooling of the bolometers to 0.1~K. It is
        shielded with a dedicated 1.8~K screen. Each bolometer can
        also be fed via an optical fibre in the back of the bolometer,
        with near infrared ligth provided by a diode which acts as an
        internal relative calibrator.  }
         \label{cryoplate}
   \end{figure}

In order that future versions of the photometer can accomodate small
arrays of bolometers on each channel, imaging cold optics have been
designed for Diabolo. Using lenses rather than mirrors, the system is
compact enough that two (and possibly three in a next version) large
throughput channels fit into a small portable dewar.
The sky is imaged through a cold pupil lens onto a cold focal plane
lens.  For each channel, the light is then fed by another lens onto
the bolometer and its associated Winston cone. The lenses are made of
quartz (of index of refraction 2.14) with anti--reflection coatings
adapted to each wavelength. As in the {\sc PRONAOS--SPM} photometer
(Lamarre \etal 1994), the optical plate is sustained below the
cryostat by three pillars and contains the optical and filtering
systems (Fig.~\ref{coldplate}). It is shielded by a 1.8~K screen
covered with eccosorb. The cryogenic plate (Fig.~\ref{cryoplate}),
which is in direct contact with the lHe cryostat (pumped to 1.8~K),
receives the dilution fridge (Section~\ref{se:dil}) which provides
cooling of the two bolometers (Section~\ref{se:de}). Ray--tracing was
done including considerations on diffraction in order to optimise the
parameters of the lenses and cones (with limited use of ASAP
software). Care was taken to underilluminate the secondary and primary
mirrors to reduce sidelobe levels (the photometer effectively uses 2
metres out of the 2.6~m of the primary mirror of the Testa Grigia
telescope).


   In inverse propagation mode, the beam exiting the photometer has a
   5.6 f ratio and the useful diameter of the exit (plane parallel
   high--density polyethylene) window is 27.5~mm.  This matches the
   bolometer throughput of $15 \zu mm^2sr$, \\
   well above the diffraction limit for both channels which is 2.3 and
   $6.4\zu mm^2sr$ in channel 1 and 2.

\subsection{Filtering system}

\begin{figure*}[t]
  \includegraphics[angle=90,width=\textwidth,origin=bl]
    {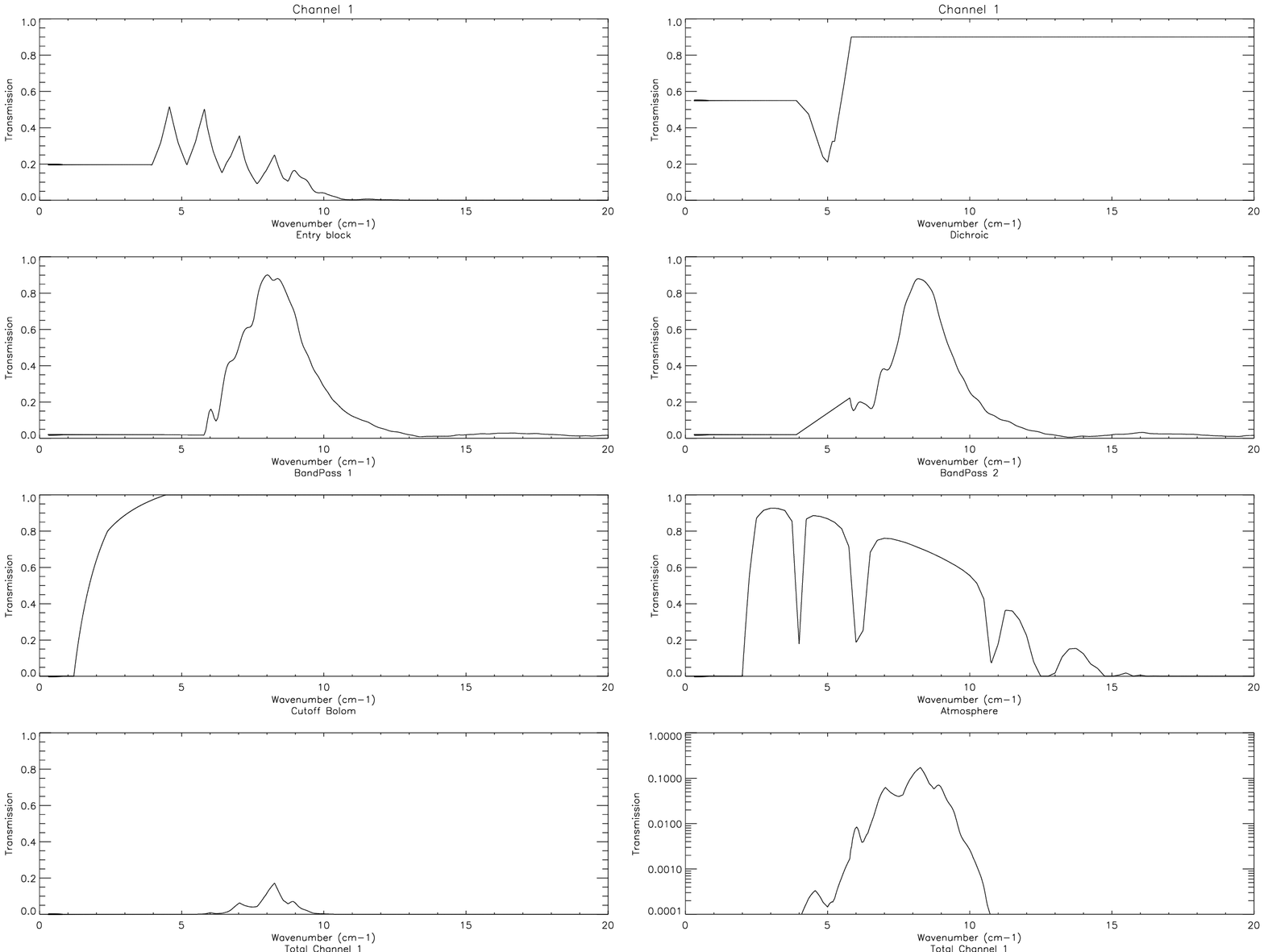}
      \caption{ Details of the filters that are used in Diabolo
        Channel 1. For each plot, the element transmission is shown as
        a function of the wavenumber. Except for the atmospheric
        transmission and the bolometer cut-off, the data are actual
        measurements interpolated onto a common grid (a constant level
        extrapolation was done to wavelengths larger than 3~mm). The
        entry block, situated next to the cold pupil, contains one
        C103 (from IRLabs) at 77~K on the penultimate cryostat screen,
        and a series of filters on the 1.8~K stage after the lens:
        another C103 filter, a diamond powdered polyethylene filter
        and 2 by 3 resonant capacitive grids (made at IAS) to insure a
        sharp submillimetre cutoff (F1 and F2 in
        Fig.~\ref{coldplate}).  The dichroic beam splitter is made of
        3 resonant capacitive grids each deposited on a mylar
        substrate and separated by 383 \micron (it was measured only
        around 5~cm$^{-1}$).  The 2 bandpass
        filters are made of free-standing metal mesh (platinum, copper
        and silver alloy, made in IKI, Moscow) of 10\micron thickness.
        The bolometer cut-off comes from diffractive effects from the
        entrance Winston cone before the bolometer (Harper \etal
        1976).  The atmospheric transmission is for 3~mm precipitable
        water vapour (cut below 2~cm$^{-1}$). The last 2 bottom figures show
        the global transmission in linear and logarithmic (resp.)
        scale, including the atmospheric transmission. }
         \label{fig:fil1}
   \end{figure*}

\begin{figure*}[t]
  \includegraphics[angle=90,width=\textwidth,origin=bl]
    {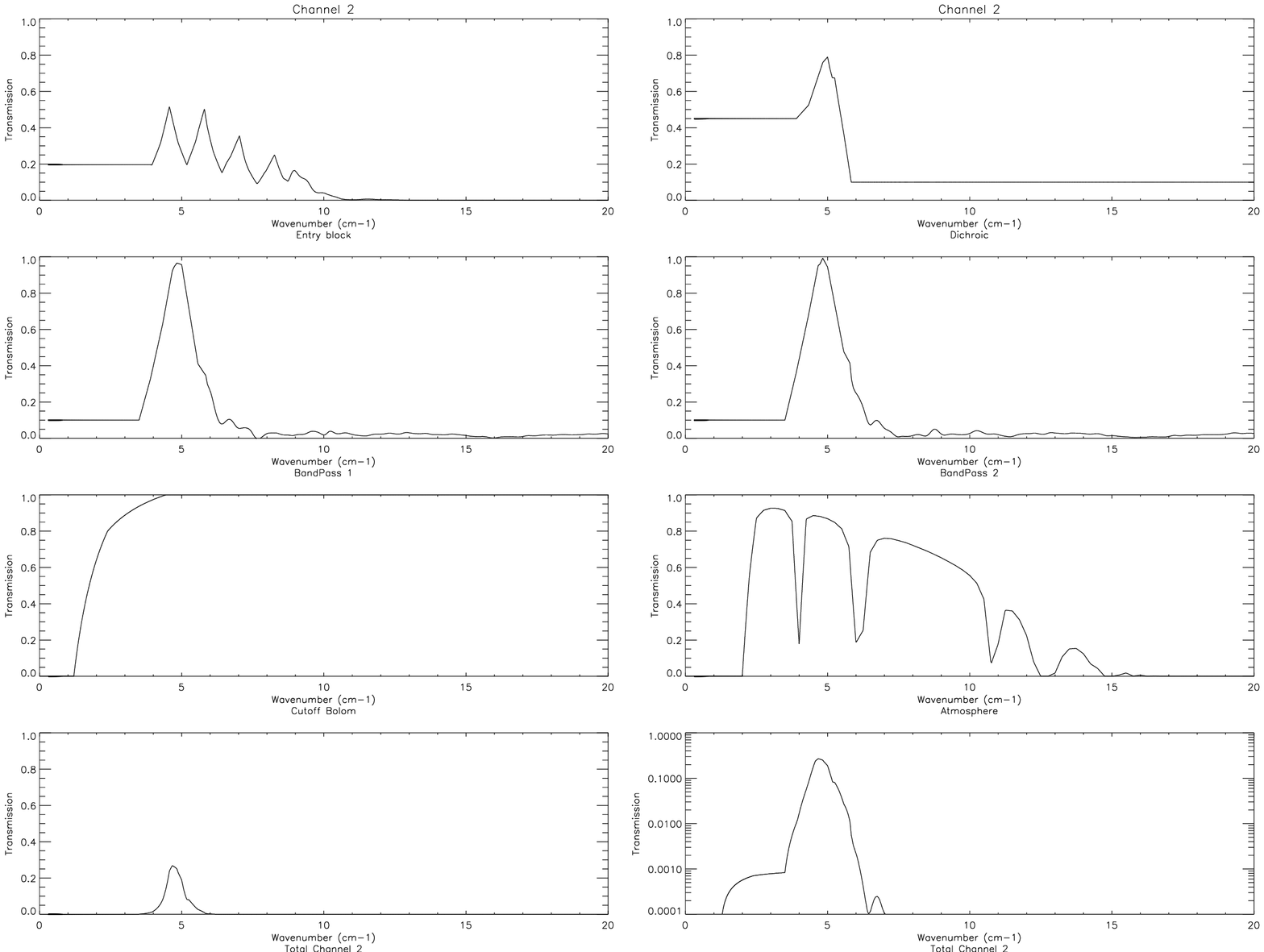}
      \caption{ Details of the filters that are used in Diabolo
        Channel 2. See previous Fig. for explanations.}
         \label{fig:fil2}
   \end{figure*}

We have devised a filtering system in order to select the appropriate
wavelengths while avoiding submillimetre radiation that would load the
bolometers. This system does not rely on the atmosphere to cut
unwanted radiation. Figures~\ref{fig:fil1} \& \ref{fig:fil2}
summarise the different filters, which are all at 1.8~K temperature
except for the first infrared cutoff filter (77~K). Measurements were
done on each element separately at room temperature only and at normal
incidence.  In the submillimetre up to 1.8~mm, this was accomplished
with a Fourier Transform Spectrometer, with a 0.3~K bolometer as the
detecting device at the Institut d'Astrophysique Spatiale (IAS)
facility.  Several measurements around 2~mm were done with a heterodyne
receiver and a carcinotron emitter at the Meudon Observatory facility
(DEMIRM).
Once all the measured transmissions are multiplied together we find an
overall expected photometer transmission which is a factor 2.5 larger
than the transmission deduced from point-source measurements. A large
fraction of the discrepancy can be attributed to the optical elements
that were not included in the calculation: the cryostat entrance
window and the lenses, as well as to some diffractive optical losses.


\section{The dilution cryostat}
\label{se:dil}

\begin{figure}[tb]
  \includegraphics[angle=0,width=\columnwidth,origin=br]
  {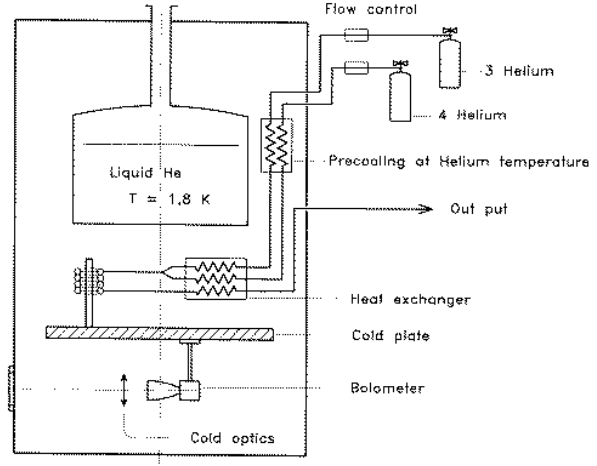}
       \caption{Schematic drawing of the Diabolo open--cycle dilution
        refrigerator. See Sect.~\ref{se:dil}. The cold plate is at
        100~mK. The lHe is pumped in order to reach a temperature of
        1.8~K which is required by the dilution fridge. }
         \label{Dilution}
   \end{figure}

In order to have a system noise as close as possible to the photon
noise, we decided to cool the detectors to 0.1~K (see
Subsect.~\ref{ss:centmilliK}). The development of a 0.1~K cooling
system fully compatible with balloon-borne and satellite environments
has been pursued at the Centre de Recherches sur les Tr\`es Basses
Temp\'eratures (CRTBT) in Grenoble (Benoit \etal 1994a, Benoit \&
Pujol 1994). The compactness and ease of use of this system render it
very attractive even for ground-based photometers.  Conversely, the
Diabolo photometer provides a good testbed for this refrigerator
before it is used on space missions.  Figure~\ref{Dilution} shows the
layout of the dilution cryostat. This new refrigerator (Benoit \etal
1994b, Sirbi \etal 1996) is the first prototype of a concept that has
become the baseline for the ESA \ps\ mission (formerly COBRAS/SAMBA).
Its principle is based on the cooling power provided at low
temperature by the dilution of $^3$He into $^4$He.  The system does
not use gravity.  Instead, the fluids are forced into room temperature
capillaries which, after going through a liquid nitrogen trap, are
thermalised by the various shields in the cryostat down to the plate
at (pumped lHe) 1.8~K.  The two Helium isotopes come from high
pressure storage vessels (see Fig~\ref{Dilution}) through flow
controllers. Typical flow rates are 3~$\mu$moles of~$^3$He per second
and 16~$\mu$moles of~$^4$He per second.
The cooling at the low temperature plate is produced by mixing the two
isotopes.  The available power is small (only few hundred nanoWatts).
Therefore the cold plate is mechanically supported by Kevlar cords
and shielded electrical wires (for the bolometers) are thermalised on
the heat exchanger (capillaries of 200 and 40~\micron diameter).  The output
mixture flows back through the heat exchanger in a third capillary
which is thermally tied to the two input capillaries. The output gas
is stored in a low pressure container for later recycling through
purification (it will be thrown away in space in case of a satellite
version).  The dilution fridge was continuously running during the
campaign (\ie for three and a half weeks), keeping the bolometers at
the useful temperature of about 0.1~K, except during the main cryostat
helium refilling, which required heating-up the cold plate temperature
to 4 K. The absolute temperature of the 0.1~K stage is measured with a
Matsushita carbon resistance ($1000\Omega$ at 0.1~K) in a AC low power
bridge.

\section{Design and calibration of the bolometers}
\label{se:de}

\subsection{Design}

\begin{figure}[tb]
  \includegraphics[angle=0,width=\columnwidth,origin=br]
  {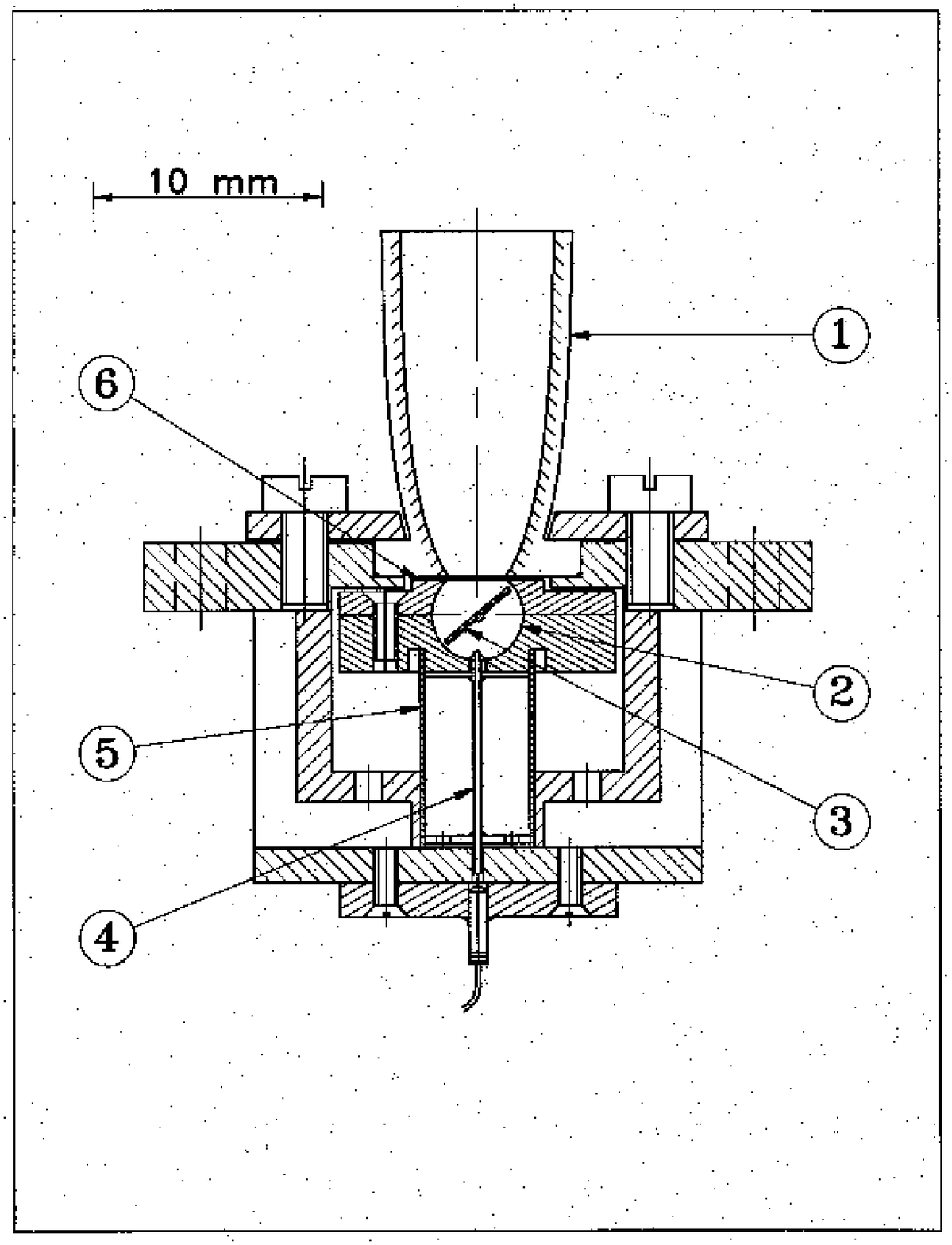}
       \caption{ 
         Schematic drawing of one bolomotric detector with its optical
         Winston cone (1) and its integration sphere (2) (4~mm
         diameter). The diamond substrate (3) is coated on the rear
         surface with a resistive bismuth layer (square impedance of
         $100\,\Omega$ to maximise absorption). It is tilted in order
         to prevent that the first reflexion goes directly out. An
         optical fiber (4) (made of a bundle of 20~\micron silica
         fibers) of 180~\micron external diameter enters the sphere,
         permitting control of the response stability at any time. A
         low pass thermal filter (5) made of a low thermal
         conductivity stainless tube (5~mm diameter; wall thickness
         250~\micron) with a cut-off at about 0.2~Hz supresses the
         noise coming from temperature fluctuations of the dilution
         cryostat. A 50~\micron gap in (6) is necessary to prevent
         short circuit of the thermal filter.  }
         \label{bolofig}
   \end{figure}

The bolometers have been developed at IAS, and benefitted from studies
in the 40~mK - 150~mK range done for thermal detection of single
events due to X--ray or $\beta$ sources (Zhou \etal 1993) or to recoil
of dark matter particles (de Bellefon \etal 1996). The design (see
Fig~\ref{bolofig}) is that of a classical composite bolometer with a
monolithic sensor as devised by Leblanc \etal (1978). The absorber is
made of a diamond window (3.5~mm diameter and 40 microns thickness)
with a bismuth resistive coating ($R = 100\, \Omega$) to match optical
vacuum impedance. The sensors were cut in a selected crystal of NTD Ge
to obtain an impedance around $10\,{\rm M}\Omega$ at 150~mK (the
effective temperature of the bolometers during these observations,
because the thermal and atmospheric backgrounds load the bolometers
above the 0.1~K cryostat temperature).  The whole sensitive system is
integrated in an integration sphere coupled to the light cone.
Moreover, by using an inclined absorber with a larger diameter than
the 2.5~mm diameter of the output of the light cone we finally
increase the optical absorption efficiency $\eta$ from 40 to 80\%,
before residual rays go out of the integration sphere (see
Eq.~\ref{eq:eta}).  Indeed, the optical efficiency can be estimated
with the well--known Gouffe's formula (Gouffe 1945), by considering
that the effective cavity surface $S$ is twice the surface of the
resistive bismuth coating (which has an emissivity larger than 0.4:
Carli \etal 1981). The entrance surface $s$ is the 2.5~mm diameter
output of the cone. With $S/s= 2 (3.5/2.5)^2=3.9$, the final cavity
emissivity is larger than 0.8.  An additional internal calibration
device (Fig~\ref{bolofig} and \ref{cryoplate}: a near infrared light
fed by a diode on the back of the bolometer via an optical fibre) was
sucessfully tested but not subsequently used, once the optics was
available.

\subsection{Calibration}

The theory of responsivity and noise from a bolometer has been written
by Mather (1984) and Coron (1976).  At the equilibrium, the Joule
power dissipated in the bolometer, $P_J$, and the absorbed radiation
power, $P_R$, are balanced by the cooling power $P_c$ due to the small
thermal link to the base temperature:

\begin{equation}
  P_c(T_1, T_0)= P_J+ P_R,
\end{equation}

with \\

\begin{equation}
  P_c= \frac{A}{L} \int^{T_1}_{T_0} \kappa(T) dT= 
  g \left( \left( {T_1\over T_g} \right) ^\alpha -
  \left( {T_0\over T_g} \right) ^\alpha \right),
\end{equation}

where $A$, $L$, and $\kappa$ are respectively the cross section,
length and thermal conductivity of the material which makes the
thermal link.  We have approximated $\kappa(T)$ with a power law,\\
$\kappa(T) \propto T^{\alpha-1}$.  From the I-V curves, we find that
for the reference temperature of $T_g= 0.1 \zu K$ the value of $g$ and
$\alpha$ are typically of $g=140$ picoWatts and $\alpha=4.5$ for both
bolometers. The impedance can be approximated with :

\begin{equation}
R(T)=R_\infty \exp((T_r/T)^\beta),
\end{equation}

where $T_r$, $R_\infty$ and $\beta$ are respectively 200K, 0.80
$\Omega$, and 0.38 for channel 1 and 20K, 52 $\Omega$, and 0.51 for
channel 2. 
The electrical responsivity at zero frequency was deduced from the I-V curves
using

\begin{equation}
S_{\rm el}(0) = \frac{Z-R}{2RI},
\end{equation}

where $Z = {\rm d}V/{\rm d}I$ is the dynamic impedance calculated at
the bias point on the I-V curve. We find electrical
responsivities of the order of $3 \times 10^7$ and $20 \times 10^7 \zu
V/W$ respectively under the sky background conditions (a load of one
to few hundred picoWatts). With a noise equivalent voltage of
typically $30 \zu nV \, Hz^{-1/2}$ above 2~Hz, the electrical NEP is
approximately of $10 \times 10^{-16}$ and $2 \times 10^{-16} \zu W
Hz^{-1/2}$ for channel 1 and 2 respectively.  The response of the
bolometer to the optical signal is linked to the electrical response
via

\begin{equation}
S_{\rm opt}= \eta S_{\rm el},
\label{eq:eta}
\end{equation}

where $\eta$ is the optical efficiency.  Thus, assuming $\eta
\ge 0.8$, the optical NEP (although not measured) at zero frequency could be reliably
estimated to be better than $15 \times 10^{-16}$ and $3 \times
10^{-16} \zu W Hz^{-1/2}$ for channel 1 and 2 respectively.

The bolometer also responds to the base plate temperature fluctuations
with a responsivity that can be deduced from the previous formalism:

\begin{equation}
{dV\over dT_0}= -S_{\rm el} {dP_c\over dT_0} = 
S_{\rm el} \frac{A}{L}\kappa(T_0) = \frac{S_{\rm el} g \alpha}{T_0}
\left( {T_0\over T_g} \right) ^\alpha
\label{eq:fluct_thermiques}
\end{equation}

The 2 bolometers that we use have typical sensitivities to the base
plate temperature of 0.2
and 1.2 $\mu {\rm V}/\mu {\rm K}$ respectively.  We see from
equation~\ref{eq:fluct_thermiques} that the larger the conduction to
the base plate and the larger the sensitivity of the bolometer to an
external signal, the most sensitive will the bolometer be to the
fluctuations of the base plate temperature. As the base plate
temperature $T_0$ fluctuates by typically $10 \zu \mu K$ over time
scales of few seconds, a regulation of this temperature should be made
in the near future to minimise fluctuations.

The time constant is less than 10 milliseconds 
for both bolometers as measured with particles absorbed by the
bolometers against a small radioactive source.

\subsection{The need for 0.1~K temperature in ground-based experiments}
\label{ss:centmilliK}

The background is relatively large in the case of ground-based
experiments. There is a general prejudice that very low temperatures are
thus not needed. Actually, the temperature required for optimised
bolometers depends only on the wavelength, because the photon and
bolometer noises both increase as the square root of the incoming background.
The general formula is (Mather 1984, Griffin 1995, Benoit 1996):

\begin{equation}
T_{\rm max}= {{hc}\over{k}} {p\over\lambda},
\end{equation}

where $hc/k= 14.4 \zu K.mm$ and $p$ is a dimensionless constant. It
turns out that for classical bolometers with a resistive thermometer,
one has typically $p\simeq 0.025$, so that the maximum temperature for
millimetre continuum astronomy is 0.4~K. Allowing for non ideal
effects and bolometers which would be 0.7 less noisy than the
background noise, a temperature of 0.1~K is required in the 2~mm
cosmological atmospheric window. The ultimate noise equivalent power
for a given background $P$ and temperature $T$ is then 

\begin{equation}
{\rm
  NEP}/({\rm W Hz^{-1/2}})\simeq 10^{-17}(P T/ (10^{-13} {\zu W K})).
\end{equation}

The present bolometers are within a factor 3 of this limit, which is
also the photon noise limit, thus leaving some margin for
improvements.

\section{Readout electronics}
\label{se:re}

\begin{figure}[tbh]
  \includegraphics[angle=0,width=\columnwidth,origin=br]
    {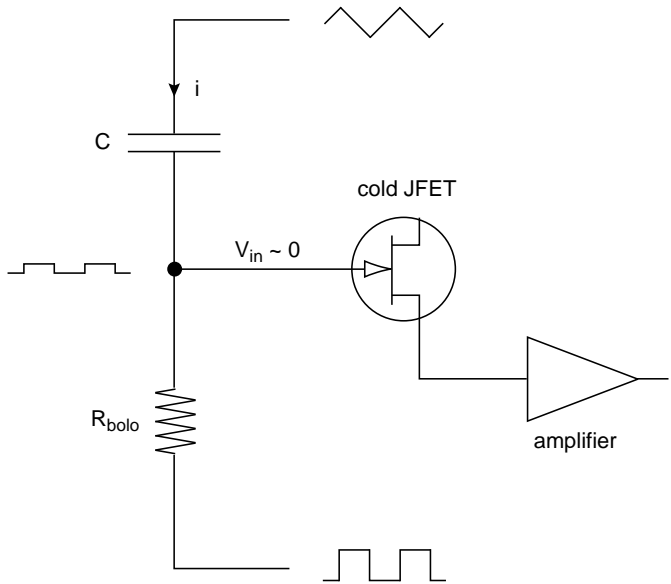}
      \caption{Principle of Diabolo readout electronics. 
        The square wave bias is adjusted to oppose the bias coming
        from the capacitive integrated current. Once these parameters are
        fixed, the out-of-equilibrium signal at the middle of the
        electrical bridge follows the varying radiation power
        absorbed by the bolometer.
        }
         \label{fig:read}
   \end{figure}

The special development made for the readout electronics is described
by Gaertner \etal (1997) in detail. Here we give the basic
characteristics of the electronics that were specially devised for
this instrument.
The bolometer is biased with a {\sl square} AC modulation at
typically 61~Hz. The current is injected through a capacitance (in
place of the classical load resistance) and an opposition voltage is
applied to ensure a near--equilibrium of the bridge (Fig~\ref{fig:read}).

Hence a small AC modulated out-of-equilibrium signal can be analysed,
which is less than $10^{-3}$ of the input voltage.  The major
advantages of this system are
\begin{itemize}
\item a constant power dissipation in the bolometer, which keeps its
  dynamical impedance constant (the square-wave signal does not
  perturb the thermal behaviour of the bolometer because it works at a
  constant input power),
\item no additional Johnson noise due to the load (which is
  capacitive rather than resistive),
\item a reduced low-frequency noise from the electronics, due to the
  modulation with a square function at frequencies above a few tens of
  Hertz.
\end{itemize}
A cold FET amplifier (JFET NJ132 at 100~K) is used to have an
amplifier noise smaller than the bolometer noise. Shielded wire is
used all the way in order to avoid electronic interferences and the
cable is soldered throughout to avoid microphonics.

Version 1 of this electronics uses an analog lock-in amplifier with a
slow feedback on the bias of the bolometer to force the signal
to be zero (with time constant of few seconds). 
In this way, we measure
\begin{itemize}
\item at intermediate frequency (1-10~Hz), the voltage variation of the
bolometer at constant current.
\item at low frequency (DC below 1~Hz), the absolute power received by
the bolometer. As the impedance of the bolometer is fixed by the
bridge balance, the bolometer works at constant temperature and the
bias power gives us directly the DC radiation input power.
\end{itemize}

\section{First observations}
\label{se:fi}

We describe now the observations done in March 1995 during a three and
a half week campaign at the Millimetre and Infrared Testa Grigia
Observatory (MITO) that gives an 8~arcminute beam with the Diabolo
photometer. Here we present some results that were acquired with a
sawtooth modulation of the secondary at 1.9~Hz (which provided a
constant elevation scan across a source, of typically 26.4~arcmin width),
combined with a slow drift of the elevation offset relative to the
source (by a total of 40~arcmin, with steps of 4~arcmin i.e. half beam
width every 10 seconds).  The acquisition frequency of 61~Hz is twice
the AC modulation readout frequency. It is synchronous with the
wobbling secondary frequency of 1.9~Hz, giving 32 measurement points
per period.

To our knowledge, these data are the first ever to be acquired on the
sky in a total power mode using unpaired bolometers. It anticipates
and proves the feasability of the total power readout mode that is
planned for next submillimetre ESA missions (\ps\ and {\sc FIRST}).

\subsection{The MITO telescope}

The MITO telescope has been specifically designed for submillimetre
continuum observations at the arcminute scale up to the degree scale,
and as such is a unique facility in Europe. The telescope (De Petris
\etal 1996), which was designed in parallel with OLIMPO (Osservatorio
nel Lontano Infrarosso Montato su Pallone Orientabile) ex TIR
(Telescopio InfraRosso) one, is a classical Cassegrain-type 2.6~m dish
with a wobbling secondary mirror designed with very low levels of
vibration (Mainella \etal 1996). The MITO facility is situated on a
dry cold site at an altitude of 3500~m close to Cervinia-Breuil in
Italy, very near the Swiss border and the Gornergrat infrared and
millimetre observatory TIRGO (Telescopio InfraRosso del GOrnergrat).
During our observations, we routinely had outside temperatures of
$-20$ Celsius (most of the nights) and good weather for about one
third of the time, making this site excellent for (sub)millimetre high
angular resolution astronomy (the opacity is less than a tenth at
zenith in the whole millimetre range).

\subsection{Data reduction}

The data were acquired with two independent acquisition systems, the
first one based on the development of the {\sc PRONAOS} one and the
second one custom made to allow for the new readout technique. The
following data analysis is based on the last system.

After deglitching, a raw map is made with the values of the signal for
given azimuth and elevation offsets. The azimuth offset is deduced
from the position in a given period of the secondary while the
elevation offset is (or should be) a sawtooth function of time.  The
registration of the instrument data with the telescope pointing
information is done with an absolute time line which happened to be
inaccurate after ten minutes of observations.  Therefore, we can only
show here the data which are post-synchronised with the help of the
occurence of a strong source detected in the raw data. The data
present a strong systematic effect which is quite reproducible and
function of the azimuth offset angle. This is easily removed from the
maps by computing the mean effect (over elevation offset angles) after
the source has been masked.  This effect is most likely due to the
instrument ``seeing'' the asymmetrical back of the secondary during
its sawtooth motion. This can and will be reduced by
adding a secondary mirror baffle as described in Gervasi \etal (1998).

Another phenomenon is the slow drift of the
detectors during time which is removed with a running constant
elevation average after the source is masked. Each map is then rotated
by the parallactic angle and coadded to the others to make a final map
in astronomical coodinates.

\begin{figure}[t]
  \includegraphics[angle=90,width=\columnwidth,origin=br]
  {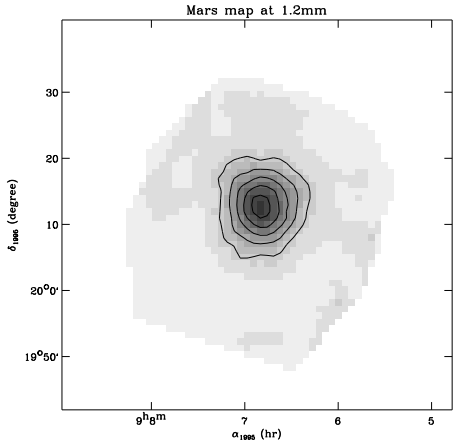}
       \caption{Mars observed with Diabolo at 1.2~mm at MITO. Contours
         are at 20, 40, 60, 80, and 100~mK$_{RJ}$ brightness levels.}
         \label{fig:Mars1}
   \end{figure}
\begin{figure}[h]
  \includegraphics[angle=90,width=\columnwidth,origin=br]
  {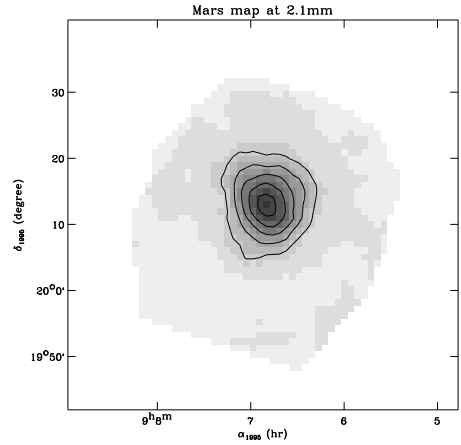}
       \caption{Mars observed with Diabolo at 2.1~mm at MITO. Contours
         are at 20, 40, 60, 80, and 100~mK$_{RJ}$ brightness levels.}
         \label{fig:Mars2}
   \end{figure}
   
   The maps of planet Mars as obtained in the two Diabolo channels,
   are shown in Fig~\ref{fig:Mars1} and \ref{fig:Mars2}. It corresponds
   to the average of 9 individual maps of $28 \times 40$ arcminutes,
   and a total integration time of 1050 seconds. The beams are quite
   similar at both wavelengths and coaligned within a precision of a
   tenth of a beam. The beam FWHM is of 7.5 arcminutes. The integrated
   beam efficiency is the same as that of an 8 arcminute FWHM Gaussian
   beam.  The signal expected from the planet after dilution in
   the beam is equivalent to a 110~mK blackbody.
   
\begin{figure}[t]
  \includegraphics[angle=90,width=\columnwidth,origin=br]
  {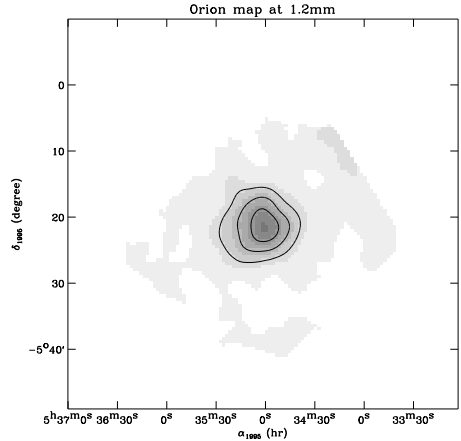}
       \caption{Orion observed with Diabolo at 1.2~mm at MITO. Contours
         are at 10, 20, 30, 40~mK$_{RJ}$ brightness levels.}
         \label{fig:Orion1}
   \end{figure}

\begin{figure}[h]
  \includegraphics[angle=90,width=\columnwidth,origin=br]
  {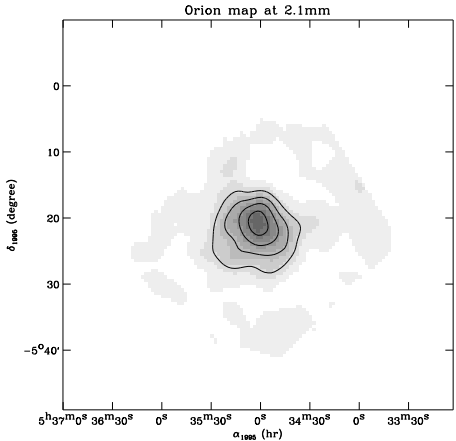}
       \caption{Orion observed with Diabolo at 2.1~mm at MITO. Contours
         are at 10, 20, 30, 40~mK$_{RJ}$ brightness levels.}
         \label{fig:Orion2}
   \end{figure}

   The Orion BN-KL nebula is detected in the raw data and the final
   maps are given in Fig~\ref{fig:Orion1} and~\ref{fig:Orion2}. It is
   calibrated with Mars signal, but no correction for differential
   extinction was applied. As Mars was at a larger elevation at the
   time of the observations, the fluxes of Orion which are found as
   $860\pm 48 \zu Jy$ and $330\pm 40\zu Jy$ at 1.2 and 2.1~mm should
   really be considered as a lower limit (especially at 1.2mm).  The
   Orion spectrum, which is dominated by dust emission in the infrared
   and submillimetre domains, clearly behaves differently at the 2.1
   millimetre wavelength, because the flux scales as the frequency to
   the power 2 between 1.2 and 2.1~mm rather than of 3 to 4 for dust
   submillimetre emission.  Free-free emission from the compact
   central HII region is most likely at the origin of the 2.1~mm
   excess.

 
   The atmospheric noise is evident in all the data that were taken.
   Sensitivities were deduced from blank sky maps as 5, 8 and 7 $\zu
   mK_{RJ}s^{1/2}$ at respectively 1.2, 2.1~mm and 2.1~mm after
   atmospheric noise decorrelation.

\section{Recent improvements}
\label{se:imp}

A number of improvements to the Diabolo instrument have been made
since the original design.  The changes made to the Diabolo setup are
listed below:

\begin{itemize}
\item All quartz lenses 
  have been replaced by polyethylene lenses, because the
  anti--reflection coatings had a tendency to fall off due to the
  stresses induced by temperature cycles.
\item A new 0.1~K cryostat has been designed with a \\
  Joule--Thomson
  cycle on the mixed Helium output, which produces the 1.8~K stage,
  hence the main lHe vessel is now at 4~K.  The major advantage is
  that refilling the cryostat with lHe is now faster because the 0.1~K
  and 1.8~K stays at the same temperature and no lHe pumping is needed
  any more. The cryogenic duty cycle of the instrument is now of half
  an hour refill every three days.
\item One bandpass filter has been removed in channel 2, in order to
  increase the sensitivity by broadening the band. We have checked
  that the small leaks that appear at high frequencies have no effect
  on the detection of the SZ effect.
\item New electronics, now fully digitally controlled with a computer
  interface, have been designed and used for subsequent observations.
  The new system is described in detail by Gaertner \etal (1997).
\item The regulation of the temperature of the thermal bath has been
  improved.  Thermometers attached to the 0.1~K stage provide
  temperature information, whereas resistances permit to heat up the
  0.1~K stage by a feedback system to stabilise the temperature in a
  closed loop.
\item Shock absorbers have been attached to the mount in order to
  minimize the microphonics induced by telescope motion (as seen by a
  general increase of the noise at all frequencies).
\item Single--bolometer detectors for each channel have been replaced
  with bolometer arrays of three bolometers in each channel.
\end{itemize}
Subsequent, upgraded versions have been used at the IRAM 30~m antenna
in Spain, and at the POM2 2.5~m telescope (without wobbling secondary
mirror) in the French Alps, in the winters 1995 through 1999, yielding
in particular, significant detections of the Sunyaev Zel'dovich effect
towards several clusters of galaxies (D\'esert \etal 1998,
Pointecouteau \etal 1999). The instrument has been open to the IRAM
community since 1998.


\acknowledgements We thank Louis d'Hendecourt and M. Gheudin for their
help in measuring the transmission of the Diabolo filters at the IAS
and DEMIRM.  We thank the Programme National de Cosmologie (ex GdR),
the INSU and the participating laboratories for their continued
support of this experiment. We also thank Pierre Encrenaz and Claudine
Laurent for their early support of the project. Part of us (M. de
Petris, P. de Bernardis, S. Masi, G. Mainella) have been supported by
Italian ASI and MURST. We thank Istituto di CosmoGeofisica (CNR) in
Turin for logistic support. Finally, we wish to thank the referee, C.R.
Cunningham, for having suggested several significant improvements
to the manuscript.
 
 
 \end{document}